# Reputation as Insurance: How Reputation Moderates Public Backlash Following a Company's Decision to Profiteer

**Danae Arroyos-Calvera**

University of Birmingham

University House, Office 1104, 116 Edgbaston Park Rd, Birmingham B15 2TY.

d.arroyoscalvera@bham.ac.uk | +0121 414 9117

**Nattavudh Powdthavee**\*

Warwick Business School

Scarman Road, Coventry, CV4 7AL

Nattavudh.powdthavee@wbs.ac.uk

---

\* Corresponding author

**Abstract**

We examine whether a company's corporate reputation gained from their CSR activities and a company leader's reputation, one that is unrelated to his or her business acumen, can impact economic action fairness appraisals. We provide experimental evidence that good corporate reputation causally buffers individuals' negative fairness judgment following the firm's decision to profiteer from an increase in the demand. Bad corporate reputation does not make the decision to profiteer as any less acceptable. However, there is evidence that individuals judge as more unfair an ill-reputed firm's decision to raise their product's price to protect against losses. Thus, our results highlight the importance of a good reputation in protecting a firm against severe negative judgments from making an economic decision that the public deems unfair.

**Citation style**: Academy of Management Journal

# INTRODUCTION

The economic impact of the COVID-19 outbreak in March 2020 has been unprecedented on a global scale. Within a few weeks of the World Health Organisation classifying COVID-19 as a pandemic, countries such as Italy, France, and the United Kingdom had shut down all of their non-essential businesses until further notice. Many companies had to cut their employees' wages or lay them off altogether to stay afloat. For example, at the time of writing, Air Canada was expected to temporarily lay off 16,000 employees because of the coronavirus crisis (Globe and Mail, 2020). A substantial increase in the demand for certain goods such as toilet paper, face masks, hand sanitiser, cleaning products, and dried foods had also led to some traders profiteering from the crisis by raising prices of their products (Sky News, 2020).

Standard economic theories view such economic decisions – e.g., laying off workers to save costs and raising prices to maximise profits from the demand increase – as rational and therefore acceptable responses to changes in the demand (Coase, 1937; Alchian, 1950; Marris, 1963). Nevertheless, some of these actions are potentially damaging to the company's reputation, which could later translate into lower profits in the future. Many empirical studies in this area have found a positive relationship between a company's reputation and their customer loyalty, customer trust, brand performance, and company value (Lacey & Kennett-Hensel, 2010; Stanaland et al., 2011; Aramburu & Pescador, 2019; Miller et al., 2020). Relatedly, several studies have also investigated whether corporate reputation protects customer loyalty and company value following an unexpected corporate crisis such as a data breach or a production shortage (e.g., Jones et al., 2000; Wei et al., 2017; Gwebu et al., 2018).

Casual observations of many of the current economic decisions in the market suggest that not all profit-seeking decisions are viewed as equally fair in the public's eyes, which could potentially have major consequences on companies' performance. For instance, although understandably angry and concerned about their financial situations, many people who got laid off amid the coronavirus crisis blamed the pandemic rather than their former employers as part of their coping strategy (BBC News, 2020). Because of the COVID-19, they see their former companies' decision to lay off workers as unfortunate but understandable. Once the pandemic subsides, it is likely that these companies will be able to resume their activity without a long-lasting damage on their corporate reputation. By contrast, the public's attitudes towards companies profiteering from the COVID-19 outbreak have been severely damaging (Financial Times, 2020). Most people appeared to view the companies' decision to profit from a demand surge as immoral, which could lead to a considerable reputation loss for these companies that might take them years to recover.

One of the first formal discussions of the importance of fairness in how individuals perceive different economic activities comes from a study by Kahneman, Knetsch, and Thaler (1986a, b). Their seminal work showed how community concerns for fairness often limit what business strategies companies could adopt to raise their profits. Through vignettes, they presented experimental evidence that individuals view the decision to raise prices to protect a company against losses (e.g., raising prices to cover the increase in the production cost) as acceptable to the point of almost being indifferent to it. Yet, the same decision to raise prices would be viewed as unfair if it was done to protect the company against losses of another source of income (e.g., raising prices due to a supply shock experienced by a competitor). Kahneman and colleagues explained their results by theorising that most individuals consider companies to be entitled to their reference profit, thus allowing them to pass on the entire loss brought about by a reduction in their profit below a positive reference level to their customers. However, customers see the exploitation of increased market power through a demand increase as unfair as such an economic decision would violate the customer's entitlement to the reference price.

Kahneman et al. (1986a, b) provide evidence that deviates from the standard economic theories, which assume that companies are subject only to legal and budget constraints when seeking to maximise profit (Coase, 1937; Alchian, 1950; Marris, 1963). This evidence has led researchers to integrate fairness into many of the subsequent economic models (e.g., Akerlof & Yellen, 1990; Kogut et al., 1996; Fehr & Schmidt, 1999) and into different management strategies across sectors and industries (Kimes & Wirtz, 2003; Wade et al., 2006; Choi & Mattila, 2007).

We now know much more about the nature of economic decisions that the general public could perceive as unfair or unjust, and how such decisions could impact the subsequent reputation and the company's performance. However, much less is understood about the opposite effect: how a company's reputation might moderate such fairness judgments. At present, there is little evidence – either correlational or causal – of the protective effect of a good corporate reputation against adverse public reactions towards a company's decision to adopt an unfair economic strategy. Could a highly reputable company carry out occasional unfair economic practices without resulting in a public backlash? Similarly, little is known about whether a bad corporate reputation negatively biases individuals' perception of an acceptable economic strategy such as a company's decision to increase prices to protect against losses. Are individuals more likely to judge an acceptable economic decision as unfair if it was carried out by a company with a poor reputation record? These questions are important if we want to build a complete picture of how reputation affects internal and external stakeholders' overall perceptions of a company.

Our aim is to provide empirical evidence of the effect of a company's reputation on how fair individuals perceive this company's economic decisions to be, and the interaction between this reputation and the unacceptable (or otherwise) nature of the decision. Like others before us (e.g., Walsh et al., 2009; Hillenbrand, Money and Ghobadian, 2013), we study the effect of CSR engagement and reputation from a stakeholder, rather than the company's, perspective.

We contribute to the corporate reputation literature by conducting a randomised vignette study to test whether a company's reputation as represented by its corporate social responsibility (CSR) records or its CEO reputation, influences how fair people judge their economic practices. Our main research question concerns the possible role of reputation as a force influencing people's fairness attributions to the company's decision. That is, the ability of a good reputation in protecting or 'shielding' a company that engages in an unfair action from a loss in reputation and business, and the ability of a bad reputation to tarnish the perception of company's fair economic action in the market. We also aim to corroborate the amplification effects found in the literature, by which a good reputation can boost the perceived fairness of an action (the 'halo' effect) and a bad reputation can make the perception of unfair economic practices even worse (the 'horn' effect).

Finally, we test how corporate reputation affects people's willingness to buy, work, and invest in the company following its decision to either profiteer from a demand surge or protect themselves against losses. Given the experimental nature of our study, we can interpret the evidence as causal effects of corporate reputation on individuals' perceptions of the company's behaviours in the market.

## RELATED LITERATURE

The business ethics literature frequently presents corporate reputation, which is the collective perception of the organisation's past actions and expectations regarding its future actions as one of the company's most valuable intangible resources (Fombrun, 1996; Fombrun & Van Riel, 1997; Walker, 2010; Chun et al., 2019). Companies are increasingly asked to visibly adopt industry-standard CSR policies as one of the corporate strategies to help build their reputation. Doing so has been empirically shown to positively influence how customers evaluate the quality of the company's products (Chernev & Blair, 2015). CSR activities have also been demonstrated to increase customer loyalty and customer trust in the company's products (Stanaland et al., 2011), employee satisfaction (Valentine & Fleischman, 2008), and company performance (Lai et al., 2010; Miller et al., 2020) through its effects on corporate reputation (Aramburu & Pescador, 2019).

There is also evidence that corporate reputation built from accumulating CSR activities protects a brand that has recently gone through an unanticipated crisis such as a human-error accident or a product-harm issue (Jones et al., 2000; Klein & Dawar, 2004; Coombs & Holladay, 2006; Wei et al., 2017; Gwebu et al., 2018; Zhang et al., 2020). However, only a few studies have investigated corporate reputation's role in moderating the negative effects of certain economic strategies on the outcomes of value to the stakeholders. One such example is a study by Williams and Barrett (2000). Using a sample of 196 companies listed on the Fortune 500 between 1991 and 1994, they show that the decline in corporate reputation associated with a company's criminal activity (e.g., a violation of the Environmental Protection Agency (EPA) and the Occupational Safety and Health Administration (OSHA) regulations) is reduced for those companies more heavily involved in corporate philanthropy. In a similar vein, Janney and Gove (2010) show that a history of CSR initiatives partially protects a company against adverse market reactions to its disclosures of involvement in the US stock option backdating scandal. In other words, these studies demonstrated that corporate reputation affects how individuals perceive companies that have engaged in illegal activity.

While there is evidence of a protective corporate reputation effect following a company's disclosure of illegal activities, most corporate decisions that individuals in the community would judge as unfair are, in fact, lawful. Empirical evidence on the interactive effect between corporate reputation and unfair economic decisions on public perceptions is scarce. As a result, we know very little whether corporate reputation protects the company against a public backlash from, for example, a decision to raise the prices of their protective equipment products during the COVID-19 pandemic, which is legal but can be judged as unfair by individuals in the community (Kahneman et al., 1986a, b). Given that most companies will have several opportunities in their lifecycle to capitalise on an increase in their market power, it seems relevant for companies to know how such economic decision will impact its long-term reputation and performances in the market.

A related question is whether corporate reputation linked to the personal reputation of the CEO would have the same effect as corporate reputation connected to the company's CSR records. Previous studies have found CEO reputation to positively impact many of the company's success indicators, including positive stock returns, company finance, and company performance (e.g., Radbourne, 2003; Anderson & Smith, 2006; Francis et al., 2008). There is also evidence that CEOs with a good reputation, which is proxied by how frequent public media or newspapers mentioned the CEO's name (Milbourn, 2003), are also simultaneously viewed as able, reliable, honourable, and attractive (Park & Berger, 2004). More recently, Weng and Chen (2017) use data from Taiwan's top 150 listed companies from 2003 to 2014 to show that CEO reputation positively impacts company performance even when the company's CSR records are lacking.

**THEORY AND DEVELOPMENT**

What explains why corporate reputation might have a protective effect on the public opinion about the company's relatively unfair economic decisions? According to signal theory, corporate reputation strategically signals the company's images to relevant stakeholders, which they then use to form the company's overall impression (Fombrun & Shanley, 1990). Once established, individuals are likely to maintain harmony among their beliefs, attitudes, and behaviours in a way that is consistent with their internalised impression of the company. Suppose the company's action generates cognitive dissonance for the individuals (e.g., committing illegal acts or adopting an unfair practice). In that case, individuals may reduce inconsistencies by rationalising the company's undesirable behaviour in a way that allows them to maintain their prior beliefs (Festinger, 1962). By contrast, individuals may judge an ill-reputed company's economic decision as worse than it objectively is because its action might seem inconsistent with the signal that individuals received about the company's image.

Legitimacy theory (Burlea & Popa, 2013) provides another alternative explanation for the phenomenon. According to this theory, a disclosure of CSR activities sends out a signal that the company has the community's best interest at the heart of its operation (Hooghiemstra, 2000). If the company does not violate the social contract set by their CSR commitment, it can continue to profit from the license to operate (Cramer, 2002). Hence, the legitimacy theory implies that individuals may rationalise or tolerate a company's decision to act unfairly in the market (e.g., raising prices to capture a rise in the demand) as long as it is not in breach of the social contract. On the other hand, ill-reputed companies may not get this benefit, making it harder for their economic activity to raise profits. Consequently, their actions may be judged much more harshly by individuals in the community than the same action carried out by companies with a good corporate reputation.

Based on these theories and previous evidence on the protective effects of CSR engagement following a corporate crisis, we formulate our main hypothesis that the perceived (un)fairness of companies' economic decisions depends on the interaction between the company's reputation and how individuals would judge the economic decision independent of the company's reputation.

We hypothesise that this interaction will operate in different directions depending on the reputation valence (i.e., whether the company's CSR record is good or poor), and whether the public perception of the economic decision would be favourable or adverse. There will be instances where the CSR record and the economic decision will be aligned (they will both be positive or negative in the eyes of the public), and in other cases they will not be aligned (one will be positive and the other will be negative). The following sub-hypotheses capture our hypothesised effects.

When the reputation and the decision are not aligned, we expect the following to hold.

H1a: A good record of CSR engagement buffers the adverse public's fairness judgement of the company's decision to engage in a given economic decision.

H1b: A poor record of CSR engagement tarnishes the favourable public's fairness judgement of the company's decision to engage in a given economic decision.

When the reputation and the decision are aligned, our conjecture is that reputation will have an amplification effect on the public's fairness appraisals. We hypothesise we will find an effect akin to the halo effect, in which a good impression created in one area tends to spill over to an unrelated area (Thorndike, 1920; Nisbett & Wilson, 1977). While the halo and horn effects could also be behind the hypothesised behaviour in H1a and H1b, we assume that these phenomena are restricted to an amplification effect rather than a reversal effect.

A good corporate reputation derived from CSR may, consequently, spill over to other unrelated areas of the company's activity. Previous studies have found, for example, evidence of a halo effect in which CSR activities unrelated to the company's products (e.g., charitable donations) improve customers' perception of product quality and brand loyalty (Chernev and Blair, 2015; Chang-Hyun and Lee, 2019). Similarly, Vo et al. (2019) demonstrate how a company's CSR engagement enhances its corporate reputation, which in turn influences word of mouth about the company on Twitter.

> H1c: A good record of CSR engagement amplifies the favourable public's fairness judgement of the company's decision to engage in a given economic decision.

Operating in the opposite direction, there is evidence of a negative halo (or horn) effect in which subordinates tend to downrate supervisors' personality traits in departments where bullying prevails (Mathisen et al., 2011). Based on this, we derive our fourth hypothesis.

> H1d: A poor record of CSR engagement amplifies the adverse public's fairness judgement of the company's decision to engage in a given economic decision.

Our second hypothesis is based on the evidence outlined above that a good CEO reputation also has a positive effect on companies' performance indicators. We expect our findings about CSR record replicate when the reputation concerns the CEO of the company, rather than the company itself.

> H2: The reputation type (CEO reputation that is unrelated to the company's image vs. corporate reputation linked to their CSR record) will not have an effect on how fair companies' economic decisions are perceived.

Finally, we also test whether corporate reputation can salvage people's willingness to buy the company's product, become an employee, and invest in the company following the company's decision to either profiteer or protect against losses.

## METHODS

We conducted a between-subjects experimental vignette study using Prolific online participant pool (https://www.prolific.co/). Respondents received £0.50 as compensation for taking part.

Each participant encountered one vignette. Each vignette had two parts: a description of the reputation type and a scenario about an economic decision they made. The reputation type varied along two dimensions: (1) CSR record, and (2) CEO personal reputation, which would allow us to test H5.

The CSR reputation descriptions are shown in Table 1. They are an adaptation of those in Baumgartner, Ernst and Fischer (2020), Coombs and Holladay (2001) and Klein and Dawar (2004). Our descriptions featured the hypothetical pharmaceutical company Pharmaio and described their CSR activities linked to their commitment to human rights and environmental protections.

[Table 1 about here]

The reputation descriptions where the agent was the company's CEO can be found in Table 2. They were developed using an online study (n = 97) where participants rated behaviours (using a 10-point scale with the extremes labelled as 'worst behaviour' and 'best behaviour'). From an initial 20-item sample of behaviours, we made our final selection of 10 such that our good and bad descriptions were be balanced in terms of the magnitude of their goodness and their badness. We had no reason to assume a different magnitude of the positive and negative reputations, hence we equalised the magnitude of the two to be able to detect such an effect (if any).

[Table 2 about here]

In an additional group, which we refer to as the 'neutral description' group, the description was omitted from the vignette. We use this group as a baseline comparator.

After the reputation description, the vignettes featured a scenario that described Pharmaio's (or Pharmaio's CEO) decision to raise the price of their hand sanitiser. There were two versions of the scenario (shown in Table 3), which varied in the reason behind this decision: in one case, to protect themselves from losses derived from a change in regulation, and in the other case to increase profits following a surge in demand for their product. According to the principles of fairness in the market that Kahneman and colleagues' (1986a) developed, the public would perceive the first version to be favourable or fair, and the second version, adverse or unfair.

[Table 3 about here]

After reading the vignette, participants were asked to rate the fairness of the economic decision that was presented on a 7-point scale ranging from "extremely unfair" (-3) to "extremely fair" (+3). To avoid the starting point of the scale affecting responses, the starting point of all scales in the study was counterbalanced: in this case, that meant that half of our respondents saw the scale starting at "extremely unfair" and the other half saw the reverse scale, starting with the "extremely fair" point. For our analyses, responses were recoded to homogenise the starting point. The exact wording of the question is an adaptation of Kahneman et al. (1986a): "Read carefully about the situation below and indicate how fair or unfair do you think the decision to raise the price of their product is". Responses from this question constitute the dependent variable for testing H1.

These three factors (reputation type, reputation valence, and public perception of the decision) give rise to a 2x2x2 factorial design (plus the 'no reputation' baseline) and 10 unique vignettes. A screenshot of a sample question can be found in the Appendix.

Then, we asked three questions to capture stakeholders' behavioural intentions, which we adapted from Sen et al.'s (2006). We used a 5-point scale ranging from "very unlikely" to "very likely". Participants reported their intention to buy hand sanitiser from Pharmaio ("If you were looking to buy hand sanitizer in the next two months, you would buy it from Pharmaio"); to work for Pharmaio ("I would very much like to work for Pharmaio"); and to invest money in it ("If I had money to invest in pharmaceutical companies, I would invest at least some of it in Pharmaio").

We asked a manipulation check question to participants in the cells concerning H1 (i.e., those in the 'company reputation condition'). They had to indicate how engaged Pharmaio was with a series of activities (community support, environmentally friendly practices, workplace environment, gender equality initiatives, and anti-homophobia campaigning), some of which had not been mentioned in the vignette (the last two in the list). We expected attentive participants to choose "n/a" or "neither engaged nor disengaged" for these activities.

At the end of the study, we asked all participants about their gender, age and yearly income, which we use as control variables.

**RESULTS**

*Participants*

We collected 1,693 responses from working-aged adults (aged 18 and over) who were born and currently reside in either the UK or the US. Of those, 60.1% ($N = 1,018$) are females, 38.8% ($N = 657$) are males, 0.65% ($N = 11$) are non-binary, and 0.42% ($N = 7$) either have a different gender or preferred not to answer this question. Approximately 21% of our sample earn less than £20,000 or $20,000 per annum, and around 30% earn £50,000 or $50,000 or more. The average age is 36.24 with a standard deviation of 13.1. There were 848 (50.1%) individuals randomly allocated to the scenario where the company raises prices to profiteer, and 845 (49.9%) individuals randomly allocated to the scenario where the company raises prices to protect against losses. Table 4 breaks down these numbers further by valence (i.e., Neutral, Good, and Bad) and reputation type (i.e., CSR and CEO reputation). We can see that the minimum and maximum numbers of observations in a cell are 147 and 189, respectively.

[Table 4 about here]

A balance test produces statistically insignificant differences in gender ($\chi^2 = 0.962$), age ($\chi^2 = 0.510$), and income ($\chi^2 = 0.948$) across all cells, which implies effective randomisation in our data collection process.

*Hypotheses Testing*

This section is organised according to the two variations presented of a company's economic decision to raise prices. To vary the public acceptability of this action (i.e., how fair respondents judge it to be), we presented two different reasons behind the price increase: protecting profit or exploiting a surge in demand. Consistent with Kahneman et al. (1986a, b), participants in the neutral group judge the decision to raise prices to protect against losses ($M = -0.45, S.E. = 0.14$) as significantly fairer than the decision to profiteer from a sharp increase in consumer demand ($M = -2.01, S.E. = 0.11$). We can reject the null hypothesis that the two means are equal; a Wilcoxon rank-sum test of the two samples produces a z-ratio of $-8.80$ ($p = 0.000$).

We investigate the interaction between the acceptability of this decision and the company's reputation valence. Later, we also examine the role of reputation type (corporate CSR record vs CEO personal reputation). We begin our analysis by testing H1a-d, which allude to the corporate reputation related to the company's engagement with CSR.

What is the effect of reputation on individuals' fairness judgement of Pharmaio's decision to raise the price of their hand sanitiser to profiteer from the rise in demand? This is the comparatively unfair economic decision, out of the two presented. H1a predicts that a good CSR record protects the firm against adverse public reactions to towards the decision to profiteer, whilst H1b predicts the reverse for a company with a poor CSR record. H1c predicts that a good CSR record further enhances the general opinion that the decision to protect against losses is more acceptable than they usually are. H1d predicts that a poor CSR record would amplify the negativity of the fairness judgement.

Figure 1, which reports the overall means (and 95% confidence intervals standard error bars) of fairness judgement ratings, shows that participants generally rate Pharmaio's decision to profiteer from the rising demand as very unfair. The average fairness rating of individuals in the neutral group is $-2.08$ ($S.E. = 0.10$) on a 7-point scale that ranges from -3 ("Extremely unfair") to 3 ("Extremely fair").

[Figure 1 about here]

We find evidence that a good CSR record reduces respondents' negative reactions towards the company's decision to profiteer by approximately 24%, thereby consistent with H1a. The average fairness rating in the good reputation treatment is $-1.58$ ($S.E. = 0.11$). Using a nonparametric Wilcoxon rank-sum test, we can reject the null hypothesis that the neutral and good reputation samples' fairness ratings are from populations with the same distribution ($z = -3.72, p < 0.001$).

There is, however, little statistical evidence that a poor CSR record has amplified the negative effect of profiteering on participants' opinions. The average fairness rating in the bad reputation treatment is $-2.01$ ($S.E. = 0.11$), which is roughly the same as the average fairness rating in the neutral group. This result is inconsistent with H1d, thus suggesting that a poor CSR record does not make the public's views towards the company's decision to profiteer any worse than the benchmark.

We now move onto the more acceptable economic decision. What is the effect of reputation on participants' fairness judgement of Pharmaio's decision to raise the price of their hand sanitiser after having to adapt their building to comply with COVID-19 safety regulations? H1b predicts that companies with a poor CSR record would see the fairness perception of their action to protect profits tarnished. H1c predicts that a good CSR record would amplify the positive fairness judgement of such action.

We uncover evidence of the tarnishing effect of bad reputation on the fairness judgement of the action, which supports H1b. Compared to the neutral group, participants in the scenario where Pharmaio is known to have a history of unacceptable working conditions and a poor environmental record rated Pharmaio's decision to raise prices to protect against losses as significantly less fair ($M = -1.01, S.E. = 0.12$; Wilcoxon $z$-ratio $= 2.94, p = 0.003$).

In line with the halo effect, the average fairness rating of Pharmaio's decision to raise prices to protect against losses when Pharmaio's reputation was good is positive ($M = 0.12, S.E. = 0.14$) and statistically significantly higher than the average fairness rating of participants in the neutral group (Wilcoxon $z$-ratio $= -2.77, p = 0.006$). The evidence thus supports H1c.

Figure 2 introduces a test of H2, which predicts that the influence of the CEO's personal reputation on how fairly the economic decisions of the company are perceived to be will go in the same hypothesised direction as that of the reputation stemming from the company's CSR activity. Overall, our results support H2. Note that we had previously investigated the influences of the CEO's personal reputation on fairness judgement across a variety of scenarios akin to those presented in Kahneman et al. (1986a, b). Overall, the current study's findings replicate those obtained in that earlier experiment whose full methodology, data, and results are available from the authors upon request.

[Figure 2 about here]

A good CEO reputation has a similar protective effect as a good CSR record against adverse public opinion regarding Pharmaio's decision to profiteer. We can reject the null hypothesis that the average fairness ratings in the neutral and good reputation groups are the same (Wilcoxon $z$-ratio $= -4.25, p < 0.001$). Similar to a poor CSR record, a bad CEO reputation does not seem to make participants' opinions towards Pharmaio's decision to profiteer any worse than the benchmark.

There is some evidence that a poor CEO reputation tarnishes the company's fair actions, but we cannot reject the null hypothesis that the two means are the same (Wilcoxon $z$-ratio $= 1.19, p = 0.236$). As in the case of CSR reputation, participants tend to rate a well-reputed CEO's decision to raise prices to protect against losses as significantly fairer than the ratings obtained in the neutral group ($M = -0.02, S.E. = 0.12$; Wilcoxon $z$-ratio $= -2.19, p = 0.028$).

Figure 2's overall evidence seems consistent with H2, which implies that CSR and CEO reputations have a similar protective property against adverse public reactions towards the company's decision to profiteer.

While Figures 1 and 2's results are interesting, it remains unclear whether the differences in these fairness attitudes would ever translate into internal and external stakeholders' real actions. To formally test this, we also asked participants at the end of the survey to report their behavioural intentions regarding their intention to buy Pharmaio's product, work for Pharmaio, and invest in Pharmaio. We present the results of these behavioural intentions by economic decision and by reputation in Figures 3A-C. Note that since CSR and CEO reputations share a statistically similar protective property, we have pooled them in this analysis.

[Figure 3 about here]

We can see from Figures 3A-C that there is a clear pattern where the protective effect of a good reputation in the scenario where Pharmaio decides to exploit increased demand spillover to the three behavioural intentions. Compared to the neutral group, participants are significantly more likely to purchase Pharmaio's hand sanitiser in the next two months, work for the company, and invest in the company. We observe the same effect of a good reputation when Pharmaio decides to raise prices to protect against losses. For example, conditioning on the decision to profiteer, we can reject the null hypothesis that the average intention to buy ratings in the neutral and good reputation groups are the same (Wilcoxon $z$-ratio $= 2.03, p = 0.042$). However, as one would expect, the average intentions to buy, work, and invest are noticeably higher when the decision to raise prices was made to protect the company against losses rather than make additional profits. Here, we can reject the null hypothesis that the average intention to buy (Wilcoxon $z$-ratio $= -8.47, p < 0.001$), intention to work (Wilcoxon z-ratio $= -4.89, p < 0.001$), and intention to invest (Wilcoxon z-ratio $= -1.85, p = 0.064$) are the same between the scenarios where Pharmaio decided to profiteer and one where they decided to raise prices to protect against losses.

On the other hand, participants are significantly less likely to buy, work, and invest in an ill-reputed company. Comparing between good and bad reputation treatments, we can reject the null hypothesis that the average intention to buy (Wilcoxon z-ratio $= -8.39, p < 0.001$), intention to work (Wilcoxon z-ratio $= -16.87, p < 0.001$), and intention to invest (Wilcoxon z-ratio $= -12.86, p < 0.001$) are the same. These deterrent effects are also generally more significant when the company decides to raise prices to profiteer rather than protect against losses. Conditioning on having a good reputation, we can reject the null hypothesis that the average intention to buy (Wilcoxon z-ratio $= -7.14, p < 0.001$), intention to work (Wilcoxon z-ratio $= -4.79, p < 0.001$), and intention to invest (Wilcoxon z-ratio $= -3.56, p < 0.001$) are the same between the scenarios where Pharmaio decided to profiteer and one where they decided to raise prices to protect against losses. Hence, we have evidence that differences in participants' fairness attitudes are reflected in their behavioural intentions towards the company.

Because participants' fairness judgements and responses to the behavioural intention questions are ordinal, one objection is that they should not be aggregated and compared across groups. Hence, we estimate an ordered probit regression model that accommodates the ordinal nature of these variables and corroborate the robustness of our findings. Our predictors combine the valence of the reputation (good or bad) and the type of reputation (personal or corporate). We use the 'no reputation' group as the reference category. We also control for individual's age, income, and gender in the regressions and estimate it separately for the "raise prices to profiteer" sample (Table 5) and the "raise prices to protect against losses" sample (Table 6); see Tables 1A and 2A in the Appendix for the regression results. Since ordered probit coefficients are not readily interpretable as marginal effects, both Tables 5 and 6 report the marginal effects of the regressors on the probability of individual reporting to be in the lowest category and the highest category of fairness and behavioural intention rating.

Looking at Tables 5 and 6, we can see that the earlier results of protective reputation effects continue to hold in multivariate regressions. They are also quantitatively sizeable as well as statistically well-determined. For example, a good corporate reputation reduces the probability that the individual will report that the company's decision to profiteer is completely unfair by 16.2 percentage points (95% C.I.: -25, -7.5), whilst a good CEO reputation reduces the same likelihood by 19.5 percentage points (95% C.I.: -28.5, -10.5). On the opposite end of the scale, a good CSR record and a good CEO reputation increase the probability that the individual will report that the company's decision to profiteer is completely fair by 1.2 (95% C.I.: 0.4, 1.9) and 1.6 percentage points (95% C.I.: 0.4, 2.8), respectively.

[Table 5 about here]

[Table 6 about here]

A good CSR record and a good CEO reputation also reduce the probability that the individual will be very unlikely to buy the company's product by 11.5 (95% C.I.: -20.5, -2.5) and 17.9 (95% C.I.: -27.6, -8.3) percentage points following its decision to profiteer. The protective reputation effects are even more significant in the intention to work regressions. Here, a good CSR record and a good CEO reputation also reduce the probability that the individual will be very unlikely to work for the company by 23.3 (95% C.I.: -31.4, -15.3) and 35.6 (95% C.I.: -43.3, -27.9) percentage points following its decision to profiteer. Furthermore, although we do not observe a statistically significant effect in the fairness regression, a bad

CSR record increases the probability that the individual will be very unlikely to buy and to work for the company following its decision to profiteer from a rising demand by 11.1 (95% C.I.: 0.9, 21.2) and 12.6 percentage points (95% C.I.: 2.9, 22.3), respectively.

We can also see strong evidence of the tarnishing effect of a bad corporate reputation in the scenario where the company decides to raise prices to protect against losses. For instance, a bad CSR record increases the probability that the individual will judge Pharmaio's decision to raise prices after having to adapt their building to comply with COVID-19 safety regulations as completely unfair by 8.1 percentage points (95% C.I.: 2.8, 13.3). It also increases the probability that the individual will be very unlikely to work and invest in Pharmaio by 22.2 (95% C.I.: 13.9, 30.6) and 19.5 percentage points (95% C.I.: 11.8, 27.2).

**DISCUSSIONS**

This study contributes to the reputation literature by experimentally investigating the possible interactive effects between the company's corporate reputation and its profit-seeking strategies on public opinion and behavioural intentions. We did so in an admittedly simple way – by examining whether different, randomly-assigned descriptions of CSR and CEO reputation moderated how participants interpreted the company's decision to either profiteer from an increase in consumer demand or protect against losses by raising the price of their product.

In line with our expectations, we find a disclosure of either good CSR or good CEO reputation to substantially reduce the adverse public reactions towards the company's decision to profiteer. We also find evidence that disclosing less than favourable CSR and CEO reputation worsens public opinion and behavioural intentions even when the company's decision to raise prices was to protect against losses, which is generally considered as a reasonable thing to do during crisis (Kahneman et al., 1986a, b). Because we randomised reputation across our participants, our research method enabled us to conclude that the observed moderating effects of disclosing reputation on individuals' opinions and behavioural intentions is causal rather than correlational.

There are important implications to our results. First, the evidence of a significant protective effect of good corporate reputation implies that strategic CSR reporting should be done not only at the company level but also at the CEO level (e.g., Blankespoor and de Haan, 2020; Du and Yu, 2020; Zhang et al., 2020). The decision to disclose CSR and CEO reputation is perhaps more relevant during an economic boom, so that when the bust inevitably follows, their corporate reputation should enable them to raise prices without causing too much public backlash. Second, the evidence on a horn effect of bad reputation suggests that CEO media visibility could also be harmful to public opinion, customer loyalty, employee engagement, and brand performance following the company's decision to protect against losses during crisis. One example of this is the negative impact that the extensive media coverage of Adam Neumann's impulsiveness and self-dealing behaviours has on WeWork's IPO prospectus (Wall Street Journals, 2019).

Like all studies in social sciences, our work is not without limitations. Our results are constrained on generality, for example, by the hypothetical scenarios and the reputation descriptions generated for the purpose of this study. Notwithstanding the usual external validity argument that all hypothetical studies like ours must face, one could also imagine that different hypothetical scenarios and descriptions of CSR and CEO reputation may elicit different fairness ratings and behavioural intentions than what have been obtained in the current study. It might also be the case that participants from different cultures (e.g., Asian, and Latin American) may hold a different view regarding what may be thought of as fair profit-seeking strategies or as having a good reputation compared to those living in the UK and the US.

However, we believe that our findings have contributed new insights into the ethics literature about the protective effects of disclosing CSR and CEO reputation on individuals' judgment and decision-making.

**Acknowledgements:** We are grateful to Jack Knetsch, Despoina Alempaki, Philip Newall, and participants at the 2019 SPUDM conference in Amsterdam for providing valuable comments to the project. **Author contributions**: Both DA-C and NP conceptualised the research questions, designed the experiment, conducted the statistical analysis, and wrote the original draft together. **Conflict of interest statement:** The authors declare no conflict of interest.

**Table 1: Corporate reputation by CSR records**

| Reputation | Description |
|---|---|
| Good | Consider a prominent pharmaceutical company called Pharmaio. Since its founding, Pharmaio has been committed to respect human rights and protect the environment. *It rates highly in the "Best Workplace" survey and* sponsors several local non-profit organisations. *By awarding study grants, the company supports numerous young talents, fosters the development of their entrepreneurial and technological skills for a successful career in the pharmaceutical industry.* As part of its sustainability strategy, *the company has already improved its environmental standards in cooperation with the local universities and plans further steps to improve its already low $CO_2$ emissions and avoid hazardous chemical substances in its value chain.* |
| Bad | Consider a prominent pharmaceutical company called Pharmaio. Since its founding, Pharmaio has been committed to respect human rights and protect the environment. The company acts as a sponsor for several local non-profit organisations. *However, it has been criticised several times for the inappropriate working conditions of its employees at sites of foreign suppliers.* As part of its sustainable strategy, *the company puts efforts into reducing $CO_2$ emissions, but the current ranking on environmental performance of 17 pharmaceutical companies published by a well-known environmental conservation NGO points out that Pharmaio holds one of the back positions due to the applied hazardous chemical substances in its value chain.* |

**Note**: italic formatting added to emphasise the differences here.

**Table 2: Corporate Reputation by CEO Personal Reputation**

| Reputation | Description |
|---|---|
| Good | Alex is known to the people in her community as *honest and trustworthy*. She *is kind to* her friends and family. She is *also very respectful of other people's opinions and is always very generous* to virtually everybody she meets. <br> Alex is the CEO of a prominent pharmaceutical company called Pharmaio. |
| Bad | Alex is known to the people in her community as *a cheater who frequently lies her way out of troubles*. She *often abuses the trust that is given to her by* her friends and family. She is *manipulative by nature, and always behaves selfishly* to virtually everybody she meets. <br> Alex is the CEO of a prominent pharmaceutical company called Pharmaio. |

**Note**: italic formatting added to emphasise the differences here.

**Table 3: Scenarios**

| Scenario | Description |
|---|---|
| Fairer | Pharmaio has been selling hand sanitiser for $2. Because of the COVID-19 pandemic, *they have had to adapt their buildings to make them comply with COVID safety regulations, so in order to protect their revenue* they have raised the price of the sanitiser to $4.50. |
| Unfairer | Pharmaio has been selling hand sanitiser for $2. Because of the COVID-19 pandemic, *there is a sharp increase in the demand for Pharmaio's hand sanitiser, so in order to increase profits* they have raised the price of the sanitiser to $4.50. |

**Note**: italic formatting added to emphasise the differences here.

**Table 4: Distribution of observations per condition**

|  | Raise price to profiteer | | Raise price to protect against losses | |
|---|---|---|---|---|
|  | CSR record | CEO reputation | CSR record | CEO reputation |
| Good | 164 | 147 | 169 | 189 |
| Bad | 176 | 176 | 151 | 160 |
| Neutral | 174 | | 165 | |

**Table 5: Marginal effects of CSR and CEO reputation on fairness judgment and different behavioural intentions following the company's decision to raise prices to profiteer**

| VARIABLES | Fairness judgement | | Intention to buy | | Intention to work | | Intention to invest | |
|---|---|---|---|---|---|---|---|---|
| | Completely unfair (=-3) | Completely fair (=3) | Very unlikely to buy (=-2) | Very likely to buy (=2) | Very unlikely to work (=-2) | Very likely to work (=2) | Very unlikely to invest (=-2) | Very likely to invest (=2) |
| Bad CSR record | -0.0243 (-0.122, 0.0739) | 0.00110 (-0.00330, 0.00550) | 0.111 (0.00971, 0.212) | -0.00471 (-0.0103, 0.000876) | 0.126 (0.0295, 0.223) | -0.00449 (-0.00892, -5.97e-05) | 0.0473 (-0.0436, 0.138) | -0.00666 (-0.0195, 0.00617) |
| Bad CEO reputation | -0.0638 (-0.161, 0.0336) | 0.00329 (-0.00213, 0.00870) | -0.0202 (-0.124, 0.0831) | 0.00123 (-0.00513, -0.00759) | 0.0196 (-0.0822, 0.121) | -0.000948 (-0.00582, 0.00393) | 0.0809 (-.00670, -0.168) | -0.0103 (-0.0222, 0.00161) |
| Good CSR record | -0.162 (-0.250, -0.0753) | 0.0118 (0.00393, 0.0198) | -0.115 (-0.205, -0.0253) | 0.00931 (0.00124, 0.0174) | -0.233 (-0.314, -0.153) | 0.0276 (0.0135, 0.0417) | -0.105 (-0.178, -0.0329) | 0.0256 (0.00679, 0.0444) |
| Good CEO reputation | -0.195 (-0.285, -0.105) | 0.0161 (0.00413, 0.0280) | -0.179 (-0.276, -0.0826) | 0.0177 (0.00499, 0.0304) | -0.356 (-0.433, -0.279) | 0.0789 (0.0468, 0.111) | -0.164 (-0.234, -0.0944) | 0.0522 (0.0262, 0.0782) |
| Control variables | Yes | Yes | Yes | Yes | Yes | Yes | Yes | Yes |

**Note:** N= 842. 95% confidence intervals are in parentheses. The marginal effects are obtained from running ordered probit regressions on the relevant outcome variables (see Table 1A in the Appendix). Control variables include age, income dummies, and gender.

**Table 6: Marginal effects of CSR and CEO reputation on fairness judgment and different behavioural intentions following the company's decision to raise prices to protect against losses**

| VARIABLES | Fairness judgement | | Intention to buy | | Intention to work | | Intention to invest | |
|---|---|---|---|---|---|---|---|---|
| | Completely unfair (=-3) | Completely fair (=3) | Very unlikely to buy (=-2) | Very likely to buy (=2) | Very unlikely to work (=-2) | Very likely to work (=2) | Very unlikely to invest (=-2) | Very likely to invest (=2) |
| Bad CSR record | 0.0805 (0.0284, 0.133) | -0.0175 (-0.0310, -0.00406) | 0.0975 (0.00758, 0.187) | -0.00577 (-0.0122, 0.000667) | 0.222 (0.139, 0.306) | -0.0186 (-0.0293, -0.00796) | 0.195 (0.118, 0.272) | -0.0249 (-0.0392, -0.0106) |
| Bad CEO reputation | 0.0235 (-0.0218, 0.0689) | -0.00716 (-0.0213, 0.00696) | 0.0248 (-0.0634, 0.113) | -0.00183 (-0.00847, 0.00481) | 0.167 (0.0808, 0.254) | -0.0162 (-0.0264, -0.00591) | 0.102 (0.0286, 0.175) | -0.0174 (-0.0317, -0.00316) |
| Good CSR record | -0.0581 (-0.0959, -0.0204) | 0.0374 (0.0115, 0.0633) | -0.146 (-0.231, -0.0618) | 0.0203 (0.00501, 0.0357) | -0.132 (-0.196, -0.0683) | 0.0386 (0.0157, 0.0615) | -0.0996 (-0.156, -0.0434) | 0.0451 (0.0161, 0.0741) |
| Good CEO reputation | -0.0486 (-0.0872, -0.0101) | 0.0278 (0.00474, 0.0509) | -0.184 (-0.260, -0.107) | 0.0303 (0.0132, 0.0474) | -0.221 (-0.278, -0.164) | 0.119 (0.0768, 0.161) | -0.135 (-0.185, -0.0848) | 0.0819 (0.0494, 0.114) |
| Control variables | Yes | Yes | Yes | Yes | Yes | Yes | Yes | Yes |

**Note:** N= 839. 95% confidence intervals are in parentheses. The marginal effects are obtained from running ordered probit regressions on the relevant outcome variables (see Table 1A in the Appendix). Control variables include age, income dummies, and gender.

**Figure 1: Mean fairness judgement of economic decisions by corporate reputation gained from different CSR records**

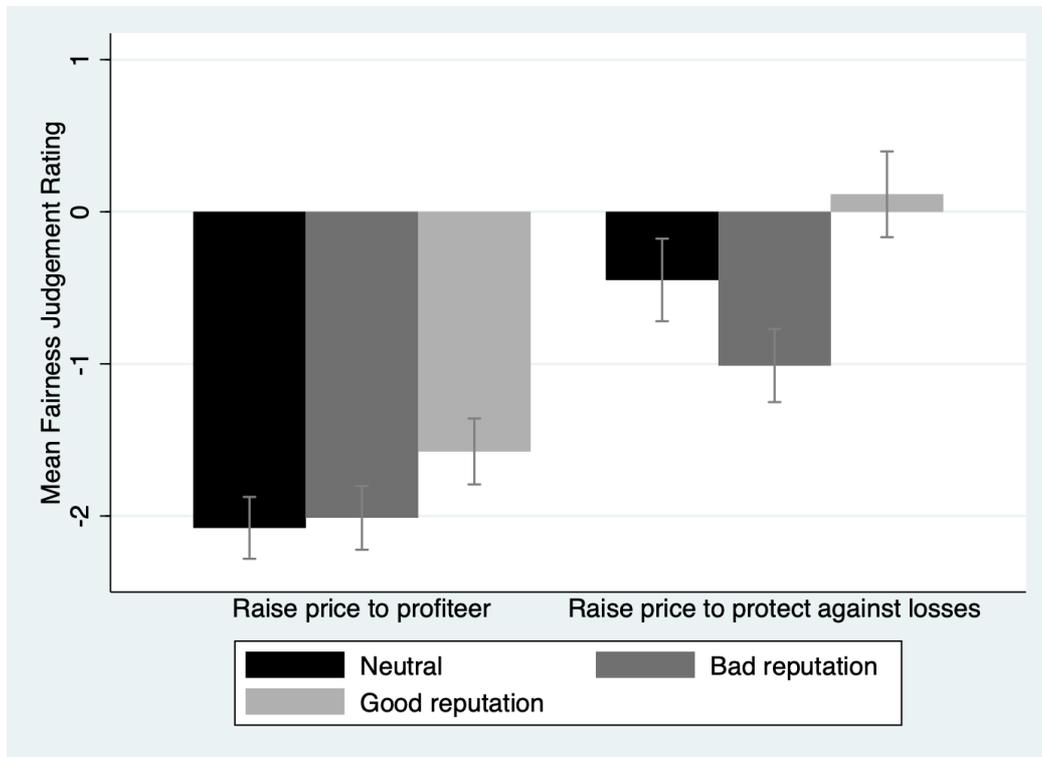

**Note:** 95% Confidence Intervals are reported.

**Figure 2: Mean fairness judgement of economic decisions by corporate reputation gained from different CEO reputations**

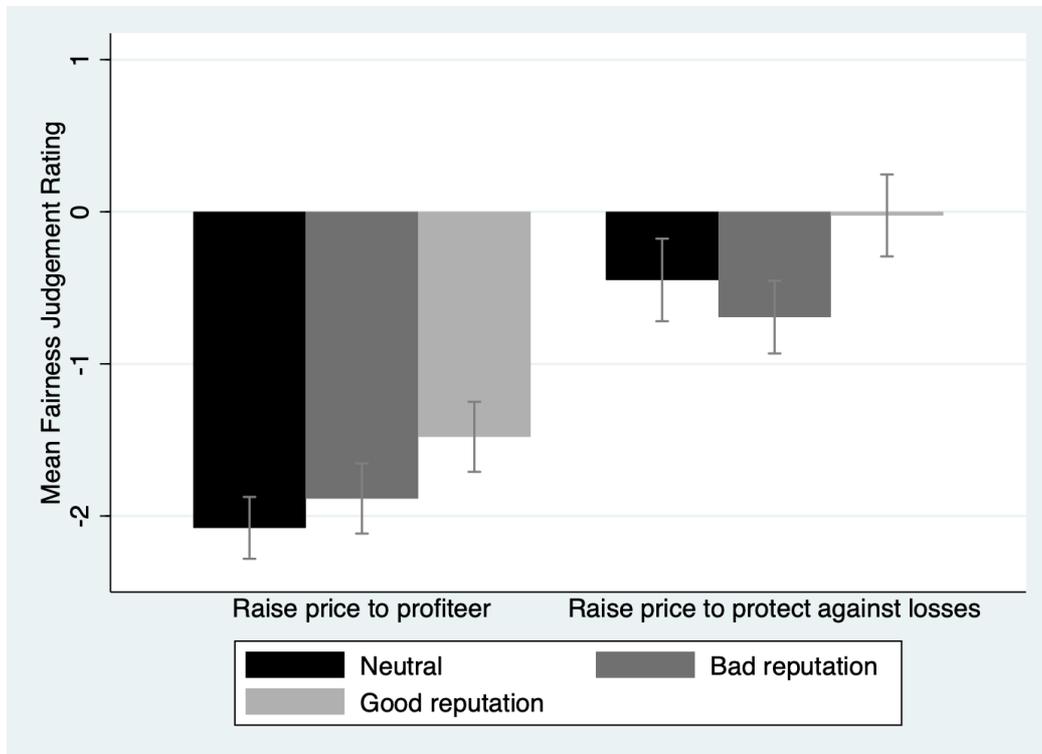

**Note:** 95% Confidence Intervals are reported.

**Figures 3A-C: Mean fairness judgment and intention to buy, work, and invest in the company: Pooled CSR and CEP reputation scenarios**

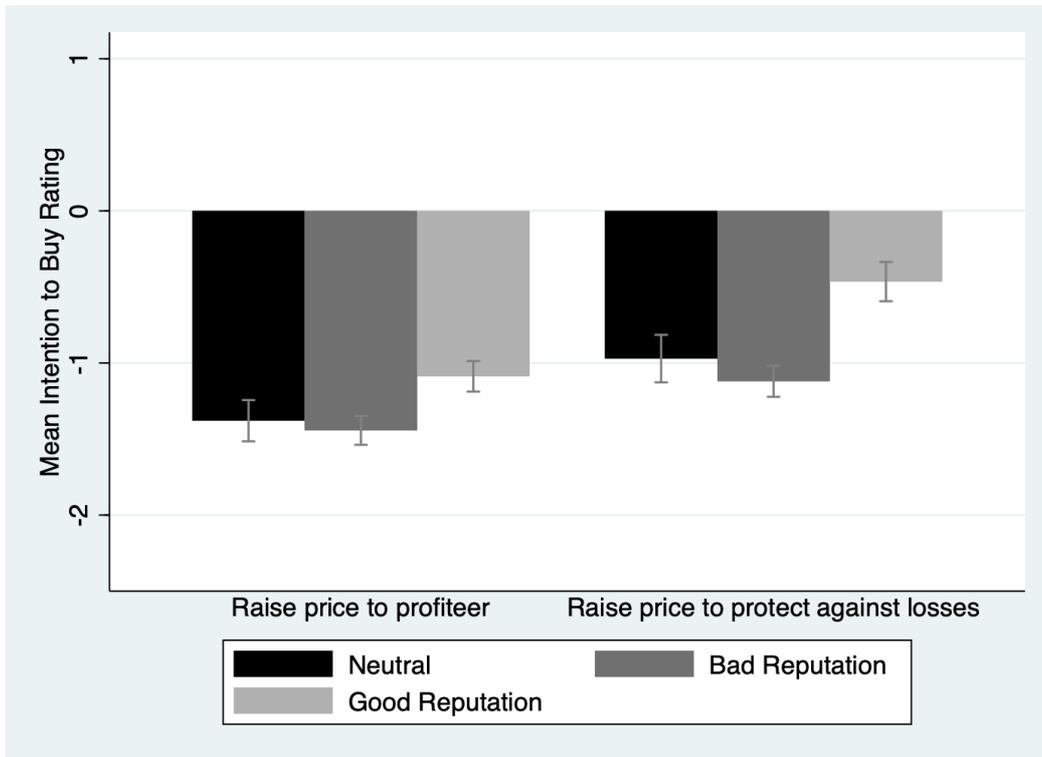

*Fig. 3A: Mean intention to buy*

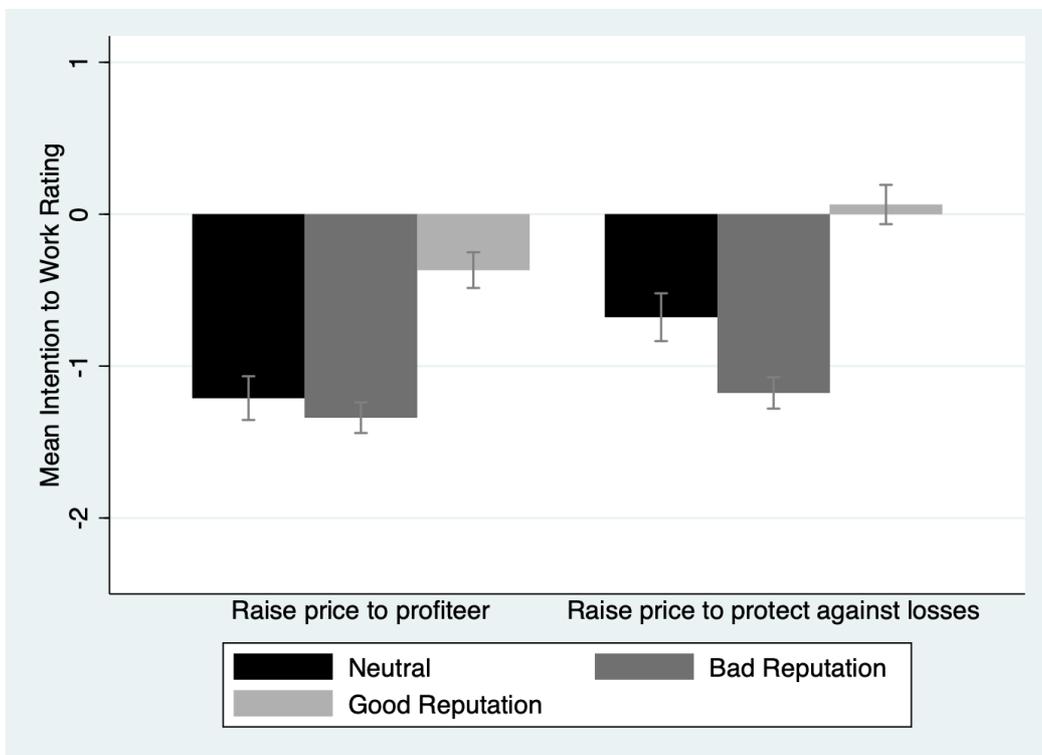

*Fig. 3B: Mean intention to work*

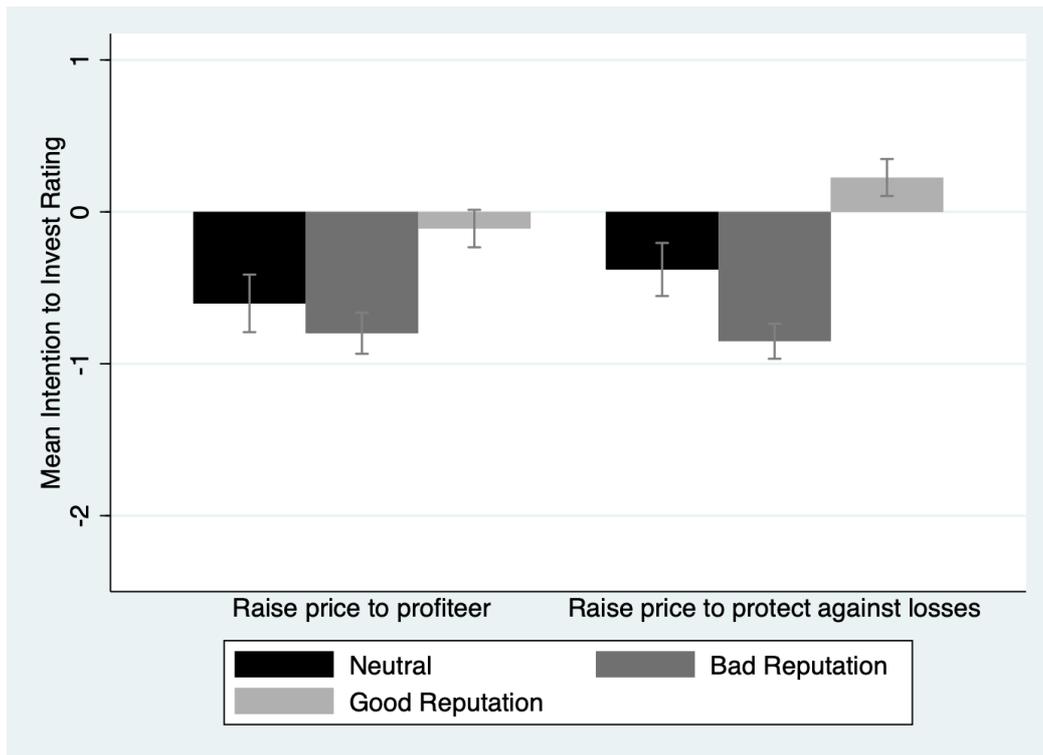

*Fig. 3C: Mean intention to invest*

**Note:** 95% Confidence Intervals are reported.

**Appendix**

## Figure A1: Sample Question

**Read carefully about the hypothetical situation below, and indicate how fair or unfair you think the decision to raise the price of their product is.**

Alex is known to the people in her community as honest and trustworthy. She is kind to her friends and family. She is also very respectful of other people's opinions and is always very generous to virtually everybody she meets. Alex is the CEO of a prominent pharmaceutical company called Pharmaio and is in charge of making all major corporate decisions.

Pharmaio has been selling hand sanitizer for $2. Because of the COVID-19 pandemic, there is a sharp increase in the demand for Pharmaio's hand sanitizer, so in order to increase profits they have raised the price of the sanitizer to $4.50.

○ Extremely fair

○ Moderately fair

○ Slightly fair

○ Neither fair nor unfair

○ Slightly unfair

○ Moderately unfair

○ Extremely unfair

**Table 1A: Ordered probit regressions used in the estimation of marginal effects in the scenario where the company decides to raise prices to profiteer**

| VARIABLES | (1) Fairness judgement | (2) Intention to buy | (3) Intention to work | (4) Intention to invest |
|---|---|---|---|---|
| Bad CSR record | 0.0631 (-0.192, 0.318) | -0.296 (-0.569, -0.0233) | -0.322 (-0.571, -0.0728) | -0.130 (-0.378, 0.119) |
| Bad CEO reputation | 0.166 (-0.0879, 0.419) | 0.0516 (-0.212, 0.315) | -0.0494 (-0.306, 0.207) | -0.219 (-0.456, 0.0189) |
| Good CSR record | 0.429 (0.197, 0.661) | 0.292 (0.0622, 0.521) | 0.628 (0.409, 0.847) | 0.323 (0.103, 0.544) |
| Good CEO reputation | 0.520 (0.276, 0.763) | 0.456 (0.206, 0.705) | 1.078 (0.834, 1.323) | 0.541 (0.314, 0.768) |
| Control variables | Yes | Yes | Yes | Yes |
| Log likelihood | -1209.63 | -941.23 | -1074.35 | -1207.68 |
| Observations | 842 | 842 | 842 | 842 |

**Note:** 95% confidence intervals are in parentheses. Possible responses to fairness judgement question range from -3. "Completely unfair" to 3. "Completely fair". Possible responses to behavioural intention questions range from -2. "Very unlikely" to 2 "Very likely". Control variables include age, income dummies, and gender.

**Table 2A: Ordered probit regressions used in the estimation of marginal effects in the scenario where the company decides to raise prices to protect against losses**

| VARIABLES | (1) Fairness judgement | (2) Intention to buy | (3) Intention to work | (4) Intention to invest |
|---|---|---|---|---|
| Bad CSR record | -0.344 (-0.567, -0.122) | -0.250 (-0.482, -0.0184) | -0.587 (-0.812, -0.362) | -0.573 (-0.803, -0.342) |
| Bad CEO reputation | -0.114 (-0.335, 0.107) | -0.0647 (-0.295, 0.165) | -0.448 (-0.681, -0.215) | -0.320 (-0.550, -0.0896) |
| Good CSR record | 0.380 (0.147, 0.613) | 0.419 (0.172, 0.665) | 0.449 (0.231, 0.667) | 0.424 (0.186, 0.663) |
| Good CEO reputation | 0.303 (0.0718, 0.534) | 0.543 (0.314, 0.772) | 0.919 (0.684, 1.154) | 0.644 (0.421, 0.866) |
| Control variables | Yes | Yes | Yes | Yes |
| Log likelihood | -1478.30 | -1124.90 | -1136.59 | -1189.81 |
| Observations | 839 | 839 | 839 | 839 |

**Note:** 95% confidence intervals are in parentheses. Possible responses to fairness judgement question range from -3. "Completely unfair" to 3. "Completely fair". Possible responses to behavioural intention questions range from -2. "Very unlikely" to 2 "Very likely". Control variables include age, income dummies, and gender.